\documentclass[prl,reprint,aps,floatfix,superscriptaddress]{revtex4-2}

\usepackage{amsmath}
\usepackage{amssymb}
\usepackage{graphicx}
\usepackage{color}
\usepackage{verbatim}
\usepackage{comment}
\usepackage{xcolor}
\usepackage{soul}
\usepackage{xr}
\usepackage{hyperref}
\usepackage{xr-hyper}
\usepackage{ulem}
\usepackage{comment}
\hypersetup{hidelinks}

\newcommand{\ba}{\begin{array}}
\newcommand{\ea}{\end{array}}

\begin{document}

\title{Nonlinear Spectral Computing}

\author{Ludovica Falsi}
\email{ludovica.falsi@uniroma1.it}

\author{Francesco Coppini}

\author{Claudio Conti}

\affiliation{Dipartimento di Fisica, Università di Roma ``La Sapienza'', 00185 Rome, Italy}

\date{\today}

\begin{abstract}
Many machine learning algorithms rely on the Fourier transform, which is also the inner core of many analog computing devices and, specifically, of optical computing. The square of the modulus of the Fourier spectrum naturally maps onto pairwise interactions, as in the renowned Ising model, making it applicable to combinatorial optimization.
Here we show how the nonlinear Fourier transform, related to integrable nonlinear partial differential equations, significantly widens
the computational capability and also enables a wide variety of computing gates.
We study specifically the case of the nonlinear Schr\"odinger equation; we introduce an input composed by ``tokens'', i.e., batches of information encoded in a multi-step potential, and we show how the nonlinear spectral computing arising in the wave propagation not only generalizes combinatorial optimization to higher order problems, 
but also realizes boolean gates. In the weakly nonlinear regime, nonlinearity generates higher-order interactions as perturbative corrections to the pairwise terms. In the solitonic regime, the discrete nonlinear spectrum performs digital computation through soliton generation. By using four input tokens, we demonstrate the programmability of all 256 Boolean functions of three variables, including 3-SAT instances. Our results show that integrable nonlinear propagation provides an unprecedented single physical platform for higher-order and logical computation, opening the way to a rich variety of new algorithms for analog computing, machine learning, and classical and quantum information processing. 
\end{abstract}

\maketitle

\noindent A common approach to combinatorial optimization (CO) is to map a problem to the energy of an Ising model, whose binary variables and interacting couplings are chosen so that low-energy configurations represent optimal or near-optimal solutions\cite{Hopfield1985,Mohseni2022}. Ising machines adopt physical hardware, where the degrees of freedom collectively evolve toward low-energy configurations, either through dynamics or controlled by algorithms such as simulated annealing\cite{Kirkpatrick1983}. Ising machines include networks of parametric oscillators\cite{Marandi2014,Inagaki2016}, spatial light modulators\cite{Kalinin2025, Pierangeli2020}, optoelectronic feedback systems\cite{Bohm2019}, nonlinear polarization oscillators\cite{Chiavazzo2025,Quinn2026}, and Kerr resonators\cite{Jin2026, Chauhan2026}. 
 
 \noindent At the core of many of these analog computers lies the equivalence between linear dispersive propagation and the Fourier transform. The Fourier transform is also the backbone of algorithms in machine learning and beyond. Analog machines implementing a Fourier transform therefore naturally tackle CO with quadratic interactions. A paradigmatic example is the spatial photonic Ising machine (SPIM) \cite{Pierangeli2019, Luo2023,Veraldi2025, Yamashita2023}, which exploits optical propagation to evaluate the Ising Hamiltonian from the low-frequency component of a spatially modulated light beam in the focal plane of a lens. A key feature is that the output of the computation is obtained at a specific spatial location. However, many CO problems involve higher-order interaction terms; their representation with "quadratization" \cite{Bybee2023,Anthony2017} introduces auxiliary variables and penalty terms. The resulting overhead increases time-to-solution, limits scalability, and imposes precision requirements on the couplings \cite{Bybee2023,Valiante2021,Nusslein2022,Mandal2020}. Direct analog computation of many-body interactions have been proposed in oscillator circuits \cite{Bybee2023} and realized in digital hardware \cite{Nikhar2024} and Kerr-soliton systems \cite{Chauhan2026}, but the higher-order terms are evaluated electronically, with an overhead that rapidly increases with interaction order\cite{Nikhar2024}.
Introducing nonlinear optical processes into SPIMs offers a route to generate higher-order interactions directly in the optical field, avoiding their electronic evaluation while retaining the intrinsic parallelism of optical propagation. However, the interaction order is fixed by the underlying nonlinearity, so that accessing a different order requires a different nonlinear process\cite{Kumar2020}. What is still lacking is an analog computing platform in which the order of the nonlinear computation can be varied within the same physical architecture.

\noindent Recently, solitonic states in the nonlinear Schr\"odinger equation (NLS) with Kerr nonlinearity have found a role in analog computing where their bistable states provide a physical encoding of Ising spins\cite{Jin2026,Chauhan2026}, with programmed interaction introduced through external optoelectronic feedback. Theoretical work \cite{Marcucci2020} has previously shown universality in neuromorphic computing with a layer formed by the NLS.
However, the computational potential of integrable soliton dynamics remains largely unexplored.  An open issue concerns the existence of algorithms driven by the nonlinear Fourier transform, which is performed spontaneously in analog propagation driven by the NLS.

\noindent In this Letter, we introduce nonlinear spectral computing based on integrable Kerr nonlinear  propagation. Integrability provides analytical access to the nonlinear evolution: by introducing a ``tokenized'' input composed by parametrized step functions, we are able to express the propagation outcome in terms of elementary functions of the input field, and we use this property to generate programmable higher-order interactions.
We demonstrate the potential of the approach in two regimes. In the weakly nonlinear evolution, nonlinear orders generate progressively higher-order spin interactions: beyond the pairwise term recovered in the linear limit, the first nonlinear correction introduces a four-spin interaction, giving direct access to higher-order Hamiltonians. In the solitonic regime, the discrete nonlinear spectrum furnishes a direct Boolean readout that we use to implement satisfiability problems. With a four-token input, i.e. a potential with four steps with parameters, amplitude and phase programming enables the access to all 256 Boolean functions of three variables. Integrable Kerr propagation thus performs spontaneously a wide variety of nonlinear computing functions programmed through the input field.
\section{Tokening Input and its Nonlinear Spectrum}
\noindent The propagation of an amplitude-modulation field $u(x,z)$ in a nonlinear dispersive medium is governed by the NLS\cite{Ablowitz_Segur}:
\begin{equation}\label{NLS}
	iu_z+u_{xx}+2|u|^2u=0;\hspace{1cm} u(x,z)\in\mathbb{C}
\end{equation}
where $x$ is the spatial coordinate and $z$ denotes the propagation distance. The NLS is a fundamental equation in nonlinear physics, whose integrable structure was first revealed through the Zakharov--Shabat (ZS) spectral problem \cite{Zakharov_Shabat}, providing the basis for its solution by the inverse scattering transform (IST). For a given input field $u(x,0)$, the ZS problem maps the initial condition onto its nonlinear spectral data (see End Matter, further details) through the scattering matrix \cite{Ablowitz_Segur,Zakharov_Shabat,Akns}:
\begin{equation}
    S(\lambda)=
    \begin{pmatrix}
        a(\lambda) & \tilde{b}(\lambda)\\
        b(\lambda) & \tilde{a}(\lambda)
    \end{pmatrix},
\end{equation}
where $\lambda$ is the spectral parameter. The terminology reflects the scattering interpretation of the ZS problem: an incident wave propagating from the asymptotic region $x\to+\infty$ through the field $u(x,t)$ is partially reflected and partially transmitted toward $x\to-\infty$. The corresponding reflection and transmission coefficients, $r(\lambda)$ and $t(\lambda)$, are expressed in terms of the scattering functions $t=a^{-1}$ and $r=b/a$ (see End Matter). The scattering matrix thus provides a complete spectral characterization of the input field, including both its radiative and solitonic components. The radiative (continuous) spectrum is characterized by the scattering data on the real axis, $\lambda\in\mathbb{R}$, whereas solitons constitute the discrete spectrum and correspond to zeros of the analytic continuation of $a(\lambda)$ into the upper complex plane $\mathbb{C}^+$. 

\begin{figure}[t]
\centering
 \includegraphics[width=0.8\columnwidth]{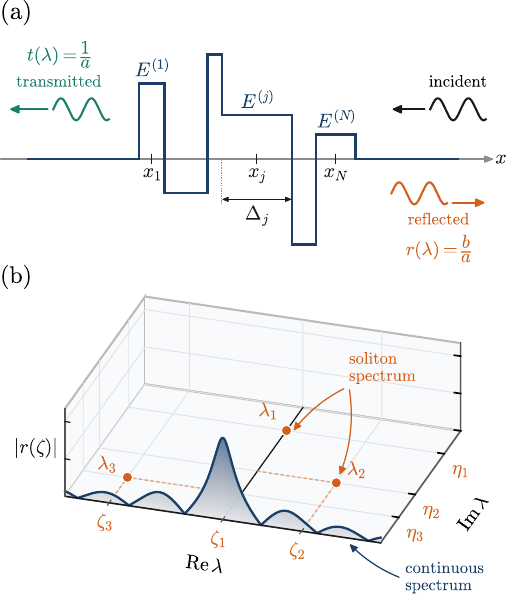}
   \caption{\textbf{Nonlinear spectral decomposition of the input field.} (a) Sketch of the scattering process for the stepwise potential in Eq.~\eqref{step_wise}. A wave of the ZS  propagating from the asymptotic region $x\to\infty$ interacts with the stepwise potential, resulting in reflected and transmitted components. (b) Associated spectrum of the initial condition. The continuous component lies on the real axis, $\lambda\in\mathbb{R}$, while the discrete component (solitons) is given by the zeros  (orange dots) of $a(\lambda)$ in the upper complex half-plane, $\mathbb{C}^{+}$:  The discrete eigenvalues are  $\lambda_j=\zeta_j+i\eta_j$ for which $a(\lambda_j)=0$ and $\eta>0$.}
    \label{fig:placeholder}
\end{figure}

We consider an input composed of $N$ rectangular function; each step has three parameters, and is referred to as a token:
\begin{equation}\label{step_wise}
	u(x,0)=\sum_{j=1}^N  \beta\left(\frac{x-x_j}{\Delta_j}\right)E^{(j)};\hspace{1cm}E^{(j)}=\sigma_j\xi_j\;;
\end{equation}
$\beta(t)$ is the box function of length and area 1, $\Delta_j$ is the length of the j-th bin, and $x_j$ its central position. The definition of $E^{(j)}$ makes explicit the spin encoding, $\sigma_j=\pm1$ and the complex amplitude $\xi_j\in\mathbb{C}$. 
The scattering matrix $S(\lambda)$ associated with the input \eqref{step_wise} is given by the ordered product of the transfer matrices of the individual token, in direct analogy with the Boffetta–Osborne algorithm~\cite{Boffetta_Osborne}:
\begin{equation}\label{scattering_N}
S(\lambda) = e^{i\lambda \frac{\Delta}{2}\boldsymbol{\sigma}_z}T_N\cdot  T_{N-1}\cdots  T_1e^{i\lambda \frac{\Delta}{2}\boldsymbol{\sigma}_z}
\end{equation}
where $\Delta=\sum_{j=1}^{N}\Delta_j$ is the total length of the input wave packet, and $T_j$ is the transfer matrix associated with the $j$th token:
{\small\begin{equation}\label{transfer_M}
	\begin{split}
T_j(\lambda)=\begin{pmatrix}
\cos(\Delta_j\mu_j)-i\lambda\frac{\sin(\Delta_j\mu_j)}{\mu_j}&&i\frac{E^{(j)}}{\mu_j}\sin(\Delta_j\mu_j)
\\i\frac{\overline{E}^{(j)}}{\mu_j}\sin(\Delta_j\mu_j)&&\cos(\Delta_j\mu_j)+i\lambda\frac{\sin(\Delta_j\mu_j)}{\mu_j}
\end{pmatrix}
\end{split}
\end{equation} }
with $\mu_j=\sqrt{\lambda^2+|E^{(j)}|^2}$  and $\overline{E}$ denoting the complex conjugation of $E$.

The transfer matrices are SL(2,$\mathbb{C}$) elements and are generally noncommutative (see \eqref{commutator}), since commutator vanishes  ($[T_i,T_j]=0$) only in these three cases: (i) the fields are identical, $E^{(i)}=E^{(j)}$, or (ii) at $\lambda=0$, when the fields have the same phase up to an overall sign, $\overline{E}^{(i)}/E^{(i)}=\overline{E}^{(j)}/E^{(j)}$ or iii) the trivial case where $\sin(\Delta_i\mu_i)=0$ or $\sin(\Delta_j\mu_j)=0$.
\subsubsection{Continuous Spectrum Computing}
\noindent When the spectrum contains only the continuous component, corresponding to sufficiently low input intensity, the transmission coefficient $t(\lambda)=1/a(\lambda)$ is directly related to the physically observable far-field intensity measured for $z\gg1$, with real values of $\lambda$ satisfying $\lambda=-x/(4z)$\cite{Zakharov_Fraunhofer}:
\begin{equation}\label{Far_field_I}
   I= |u(x,z)|^2\sim -\frac{1}{2\pi z}\log \bigg|a\left(-\frac{x}{4 z}\right)\bigg|
\end{equation}
For $\lambda\in\mathbb{R}$, the transfer matrix $T_j\in SU(2)$ therefore reduces to:
{\small\begin{equation*}
    T_j=e^{i(\boldsymbol{\nu}_j\cdot\boldsymbol{\hat{\sigma}})}.
\end{equation*}}
where $\boldsymbol{\hat{\sigma}}=(\sigma_x,\sigma_y,\sigma_z)$ denotes the Pauli matrices, and $\boldsymbol{\nu}_j$ denotes the coordinates:
\[
\boldsymbol{\nu}_j=\Delta_j 
\left(E_r^{(j)},-E_i^{(j)},-\lambda\right)^T.
\]

The scattering matrix $S$ therefore represents an ordered sequence of $N$ $SU(2)$ rotations, each determined by the physical parameters of the corresponding bin. As a first application of Eq.~\eqref{scattering_N}, we consider the commuting case. This provides the simplest nonlinear generalization of the SPIM architecture and recovers its linear limit through an expansion at low input intensity.  Without loss of generality, we set this common phase to zero, such that $E^{(j)}\in\mathbb{R}$ and
{\small\begin{equation}
\begin{split}
 	S(0) =e^{i\sum_j\Delta_j E^{(j)}\boldsymbol {\sigma}_x}.
    \end{split}
\end{equation}}
 For low input intensities, $|E^{(j)}|\ll1$, we write $E^{(j)}=\epsilon\sigma_j\xi_j$, with $\epsilon\ll1$ the perturbative parameter. Using Eq.~\eqref{Far_field_I}, the far-field intensity measured at the center of the wave train, corresponding to $x=0$ (or $\lambda=0$), is:
\begin{equation}
 \begin{split}
  I& \propto \epsilon^2J^{(0)} +\epsilon^2 \sum_{m<n}J^{(2)}_{mn}
    \sigma_m\sigma_n+\\
    &+\epsilon^4\sum_{m<n<l<k}J^{(4)}_{mnlk}
    \sigma_m\sigma_n\sigma_l\sigma_k+O(\epsilon^6)\end{split}
\end{equation}
with
\begin{equation*}
    \begin{split}
  & J^{(4)}_{mnlk}= 4 \,\tilde{\xi}_m\tilde{\xi}_n\tilde{\xi}_l\tilde{\xi}_k;\\[2mm]
  & J^{(2)}_{mn}=2 \tilde{\xi}_m\tilde{\xi}_n\left[1+\epsilon^2 \left( \sum_l\tilde{\xi}_l^2+\frac{1}{3}(\tilde{\xi}_m^2+\tilde{\xi}_n^2)\right)\right];\\[2mm]
  & J^{(0)}= \sum_l\tilde{\xi}_l^2 +\epsilon^2  \left[ \frac{1}{6}\sum_l\tilde{\xi}_l^4 +\sum_{m<n}\tilde{\xi}_m^2\tilde{\xi}_n^2 \right];
   \end{split}
\end{equation*}
where $\tilde{\xi}_j=\xi_j\Delta_j$ is the ``area'' of the bin. The $O(\epsilon^4)$ correction to the perturbative expansion introduces a new rank-one tensor $J^{(4)}$, together with corrections to the $J^{(2)}$ and $J^{(0)}$ terms.

\subsubsection{Discrete Spectrum Computing}
\noindent At higher input intensities, a discrete component emerges in the nonlinear spectrum. This component is associated with solitons, i.e., localized, nondispersive solutions of the NLS, each of them associated with its discrete eigenvalues $\lambda_j=\zeta_j+i\eta_j$, $\eta_j>0$, defined by the zeros $a(\lambda_j)=0$ in the upper complex half-plane. In the far field, $z\gg1$, an isolated soliton takes the asymptotic form
\begin{equation}
u^{s}_j(x,z)\sim
\frac{2\eta_j\,e^{-2i[\zeta_jx+2(\zeta_j^2-\eta_j^2)z]+i\psi_j}}
{\cosh[2\eta_j(x-x_j+4\zeta_j z)]},
\end{equation}
with peak intensity $4\eta_j^2$ and velocity $-4\zeta_j$, while $(x_j,\psi_j)$ are determined by the initial condition.

\begin{figure}[t]
\centering
 \includegraphics[width=\columnwidth]{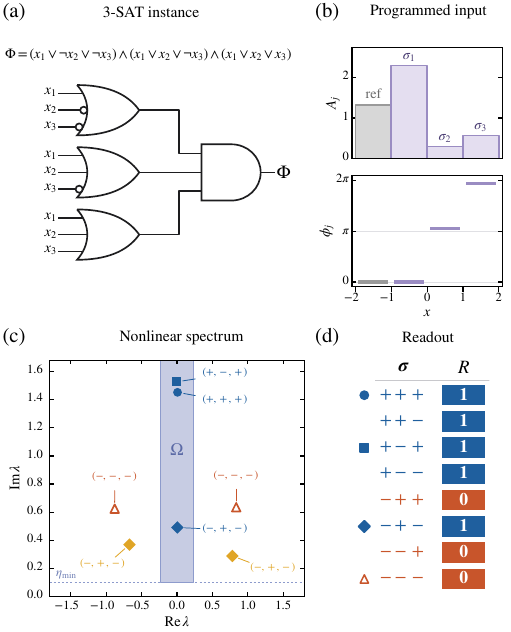}
   \caption{\textbf{3-SAT computing through the nonlinear spectrum.}
(a) Example of a 3-SAT instance with three Boolean variables and three clauses, each containing three literals; open circles indicate negations. (b) Four-token encoding of the Boolean function $\Phi$. Amplitudes $A_j$ (top) and phases $\phi_j$ (bottom) of the three variable tokens $\sigma_j$ and a reference token define the optical input. (c) Nonlinear spectra for four representative configurations $\boldsymbol{\sigma}=(\sigma_1,\sigma_2,\sigma_3)$ among the eight possible ones. Markers denote the zeroes of $a(\lambda)$ in \eqref{a_4tokens}; the shaded region $\Omega$ defines the Boolean readout, with $\eta_{\Omega}$ setting the minimum soliton amplitude. (d) Boolean readout: an eigenvalue inside $\Omega$ yields $R=1$ (satisfied); otherwise $R=0$ (violated). The zeroes are calculated using standard Newton method and validated by Cauchy argument principle \cite{Rudin}.
}
\label{fig:3SAT}
\end{figure}

Figure~\ref{fig:3SAT}(a) shows an example of a 3-SAT problem, in which a Boolean formula is expressed as a conjunction of clauses, each containing three literals. The problem consists in finding an assignment of the Boolean variables that simultaneously satisfies all clauses. We use this example to illustrate computation through the discrete spectrum. The minimal input supporting interactions beyond pairwise order consists of four bins. For $N=4$

\begin{equation}
a(\lambda)=A^{(0)}(\lambda)
+\sum_{1\le i<j\le4}A^{(2)}_{ij}(\lambda)\sigma_i\sigma_j
+A^{(4)}(\lambda)\sigma_1\sigma_2\sigma_3\sigma_4,
\label{a_4tokens}
\end{equation}
where the coefficients $A^{(0)}$, $A^{(2)}_{ij}$, and $A^{(4)}$ are controlled by the token parameters (see End Matter).

As shown in Fig.~\ref{fig:3SAT}(b), we fix one token  as a reference, $\sigma_0=1$, and encode the three Boolean variables in the remaining tokens, $\sigma_{1,2,3}=\pm1$, . Their amplitudes and phases are programmed so that the resulting nonlinear spectrum implements the desired three-variable Boolean function. As shown in Fig.~\ref{fig:3SAT}(c), the resulting discrete eigenvalues lie at different positions in the complex $\lambda$ plane depending on the input configuration. We define the binary output by the presence or absence of at least one eigenvalue within the spectral window $\Omega$ defined as the section of the upper complex plane
 $\Omega=\{\lambda=\zeta+i\eta| \;\eta>\eta_{min},\; |\zeta|<\zeta_{\Omega}\}$  with chosen values ($\zeta_{\Omega}=0.239,\eta_{min}=0.1$).

\begin{figure}[t]
\centering
 \includegraphics[width=\columnwidth]{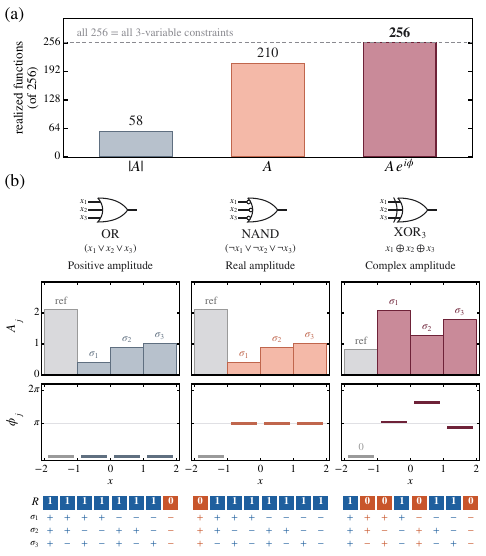}
   \caption{\textbf{Programmability of three-variable Boolean functions.}
(a) Number of accessible Boolean functions for three levels of four-token input control: amplitude only, amplitude and sign, and arbitrary amplitude and phase. Each bar gives the corresponding subset of the 256 possible truth tables. (b) Representative SAT problems requiring increasing input control. Amplitudes $A_j$ (top) and phases $\phi_j$ (bottom) are shown for the four tokens. From left to right: amplitude only, amplitude and sign, and arbitrary amplitude and phase. The reference token is fixed at $\sigma_0=1$, while the other three encode the Boolean variables.}
    \label{fig:modulazione}
\end{figure}

\begin{figure*}[t]
\centering
 \includegraphics[width=\textwidth]{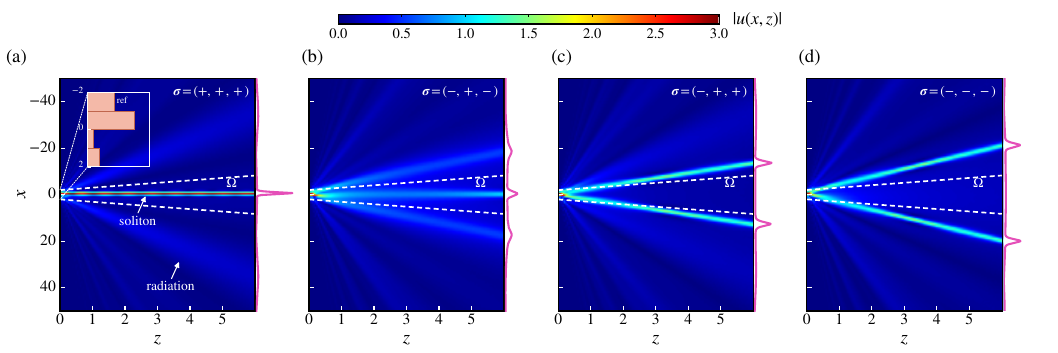}
   \caption{\textbf{Soliton readout in nonlinear propagation.} (a)–(d) Numerical propagation of representative four-token spin configurations, computed using the split-step Fourier method . The color maps show the  evolution of $|u(x,z)|$, with the localized solitonic and dispersive components indicated in (a). Dashed white lines delimit the spatial detection region corresponding to the spectral window $\Omega$. The inset shows the amplitude modulation profiles $A_j$, common to all the configurations. A discrete eigenvalue within $\Omega$ gives rise to a soliton propagating inside this region, while the remaining radiation disperses during propagation. Magenta curves on the right show the output intensity profiles at $z=z_f=6$, from which the presence or absence of a soliton within the detection region provides the Boolean readout. This provides the physical implementation of the boolean function (3SAT problem) reported in Fig.\ref{fig:3SAT}, whose architecture can be exported in all the 256 boolean functions. The numerical simulations are obtained using a refined split-step method\cite{splitstep4}}
    \label{fig:split}
\end{figure*}
Because the soliton position evolves as $x\simeq x_j-4\zeta_j z$, this spectral criterion maps directly onto a spatial detection window in the far field,
\begin{equation}
|x|<4\zeta_{\Omega} z+\frac{\Delta}{2},
\end{equation}
where the term $\Delta/2$ allows the initial soliton position to vary arbitrarily within the input pulse extension $|x_j|\leq\Delta/2$. Figure~\ref{fig:3SAT}(d) shows the resulting truth table: for each of the eight input configurations $(\sigma_1,\sigma_2,\sigma_3)$, the presence or absence of a soliton within the detection window provides the binary output of the programmed Boolean function. Since one of the four tokens is kept fixed as a reference, the remaining three tokens encode three Boolean variables, yielding eight possible spin configurations and thus defining $2^{8}=256$ possible Boolean functions. Figure~\ref{fig:modulazione}(a) compares the functions accessible with three levels of input control: amplitude-only modulation, amplitude and sign modulation, and full amplitude and phase modulation. 
Figure~\ref{fig:modulazione}(b) illustrates representative SAT problems and their mapping onto the four-token  input: the first is implemented by amplitude modulation alone, the second requires both amplitude and sign modulation, while the third requires full amplitude and phase control. The values of $(A_j,\phi_j)$ used to realize each of the 256 functions are reported in Supplemental Material.
Increasing the degrees of freedom progressively enlarges the set of realizable functions, with full complex-amplitude control, $\xi_j=A_j e^{i\phi_j}$, providing access to all 256 functions.
Figure~\ref{fig:split} shows how the spectral prediction translates into the nonlinear propagation dynamics, obtained by direct numerical integration of the NLS using the split-step Fourier method. Panels (a)–(d) show the evolution of the four-token input for representative spin configurations (the complete set is reported in the Supplemental Material), with the corresponding input profiles shown in the insets. The dashed white lines delimit the spatial detection region associated with the spectral window $\Omega$. As highlighted in Fig.~\ref{fig:split}(a), nonlinear propagation separates the localized solitonic component from the dispersive radiation. When a discrete eigenvalue lies within $\Omega$, a localized soliton emerges from the dispersive radiation and propagates inside the detection region, whereas configurations with no eigenvalue in $\Omega$ exhibit only radiation. The output intensity profiles at the propagation distance $z=z_f$, shown on the right of each panel, therefore provide a direct physical readout of the Boolean output.
Spectral points lying close to the boundary of the readout window $\Omega$, whether inside or outside it, could (in principle) compromise the reliability of the readout: perturbations in the physical setup can displace the corresponding soliton across the boundary and produce an incorrect output. To quantify this effect, we define the spectral margin $M$ as the minimum distance, in the complex $\lambda$ plane, between the discrete eigenvalues closest to the boundary of $\Omega$, and the boundary itself. For each target Boolean function, we therefore determine input modulations that not only reproduce the desired truth table but also maximize $M$. The resulting modulation parameters and the corresponding values of $M$ are listed in the table in Supplemental Material.

\section{Conclusion}
\noindent In conclusion, we have introduced nonlinear spectral computing, by using the integrable dynamics of the NLS as a computational resource. In the weakly nonlinear regime, the continuous spectrum recovers pairwise interactions in the linear limit, while successive nonlinear orders generate progressively higher-order spin interactions, with the first nonlinear correction introducing a four-spin term. In the solitonic regime, the discrete spectrum provides a Boolean readout through the presence or absence of solitons within a prescribed spectral window. For a four-token input, with one reference and three Boolean variables, programming the input amplitudes and phases gives access to all 256 three-variable Boolean functions, with direct NLS propagation translating the spectral output into a spatially resolved soliton signal.
The present three-variable implementation also provides a route toward larger constraint-satisfaction problems. Multiple clauses can be encoded in separate four-token  blocks and evaluated in parallel, each providing its own soliton readout. Alternatively, clauses involving the same three variables can be combined into a single Boolean function and directly programmed through the input amplitudes and phases, reducing the number of optical blocks. This approach can also be extended beyond 3-SAT to higher-order constraint problems, such as 3-Regular 3-XORSAT (3R3X) \cite{Nikhar2024,Kowalsky2022,Bernaschi2021}. More generally, nonlinear spectral computing provides access to a hierarchy of interaction orders within the same physical dynamics, rather than associating each computational order with a different nonlinear process. 
A distinctive feature of this framework is that different computations occur at different spatial positions (or equivalently in momentum space) as solitons appear upon evolution. This circumstance suggests that one can perform different computations simultaneously and, by choosing the reading time, select the target result; for example, at the focal plane one has the Fourier transform, at longer distances one has an Ising model with four-body interactions, and, when solitons emerge from the continuous background, the nonlinear spectrum provides the answer to a k-SAT problem. Nonlinear spectral computing is not limited to NLS and Kerr nonlinearity; it extends to all known integrable models. Therefore it opens a route toward single nonlinear architectures capable of performing a wide variety of computations with controllable higher-order interactions; many new algorithms can be conceived by 
exploiting integrability, thus opening the way to several classical and quantum applications.
\\

{\bf Acknowledgments}
F.C. and C.C acknowledge support HORIZON-ERC-2023-ADG HYPERSPIM ‘Optoelectronic and all-optical hyperspin machines for large-scale computing’ no. 101139828. L.F. acknowledges support Sapienza-Ricerca di Ateneo Projects.
Views and opinions expressed are however those of the authors only and do not necessarily reflect those of the EU or the ERC. Neither the EU nor the granting authority can be held responsible for them.

\bibliographystyle{apsrev4-2}
\bibliography{bibliography}
 \newpage 
 
\section{End Matter}

\appendix
\setcounter{equation}{0}
\renewcommand{\theequation}{A.\arabic{equation}}

The integrability of the NLS \eqref{NLS} was first discovered by associating with the NLS the Zakharov-Shabat system \cite{Ablowitz_Segur,Zakharov_Shabat,Akns}:
\begin{equation}\label{Lax_op_1}
v^{(1)}_x =- i\lambda v^{(1)}+iuv^{(2)}; \hspace{1cm} v^{(2)}_x = i\lambda v^{(2)}+i\overline{u}v^{(1)};
\end{equation}
where $\overline{u}$ denotes complex conjugation. For localized initial conditions $u(x,0)\rightarrow 0$ for $|x|\rightarrow \infty$ the standard solution technique is the Inverse Scattering Transform (IST).
The initial condition is associated with the spectral data of the operator \eqref{Lax_op_1} of the Lax pair.
We define the Jost eigenfunctions $( \boldsymbol{\psi}^{(1)}, \boldsymbol{\psi}^{(2)})$ and $( \boldsymbol{\phi}^{(1)}, \boldsymbol{\phi}^{(2)})$, which in the asymptotic region $x\rightarrow\infty$ tend to:
{\small\begin{equation*}
    \boldsymbol{\psi}^{(1)}(x;\lambda)\xrightarrow[x \to \infty]{}\begin{pmatrix}
       e^{-i\lambda x}\\0
    \end{pmatrix};\hspace{1cm}     \boldsymbol{\psi}^{(2)}(x;\lambda)\xrightarrow[x \to \infty]{}\begin{pmatrix}
     0\\  e^{i\lambda x}
    \end{pmatrix},
\end{equation*}}
and analogously $( \boldsymbol{\phi}^{(1)}, \boldsymbol{\phi}^{(2)})$ for $x\rightarrow-\infty$:\\
\begin{equation*}
    \boldsymbol{\phi}^{(1)}(x;\lambda)\xrightarrow[x \to -\infty]{}\begin{pmatrix}
       e^{-i\lambda x}\\0
    \end{pmatrix};\hspace{.5cm}     \boldsymbol{\phi}^{(2)}(x;\lambda)\xrightarrow[x \to -\infty]{}\begin{pmatrix}
     0\\  e^{i\lambda x}
    \end{pmatrix}.
\end{equation*}
The fundamental properties of the solution are extracted from the scattering matrix $S(\lambda)$, which denotes the transformation matrix between the two Jost bases:\\
\begin{equation*}
    \begin{pmatrix}
        \phi^{(1)}_1&& \phi^{(2)}_1\\\phi^{(1)}_2&&\phi^{(2)}_2
    \end{pmatrix}= \begin{pmatrix}
        \psi^{(1)}_1&& \psi^{(2)}_1\\\psi^{(1)}_2&&\psi^{(2)}_2
    \end{pmatrix}\cdot S(\lambda).
\end{equation*}
We have explicitly indicated the components of the Jost vectors $ \boldsymbol{\psi}^{(j)}=(\psi^{(j)}_1,\psi^{(j)}_2)^T$, $ \boldsymbol{\phi}^{(j)}=(\phi^{(j)}_1,\phi^{(j)}_2)^T$, for $j=1,2$. The elements of the matrix $S(\lambda)$ are called scattering functions:
\begin{equation*} 
     S(\lambda)=\begin{pmatrix}
        a(\lambda)&&\Tilde{b}(\lambda)\\[1mm]
        b(\lambda)&&\Tilde{a}(\lambda)
    \end{pmatrix},
\end{equation*}\\
The functions $a(\lambda)$ and $b(\lambda)$ are related to the reflection and transmission coefficients of a wave incident on the potential $u(x,z)$ from the region $x\rightarrow\infty$, as can be seen by analyzing the asymptotic states of $  \boldsymbol{\phi}^{(1)}$ :
\begin{equation*}
    \boldsymbol{\phi}^{(1)} \xrightarrow[x \to -\infty]{}\begin{pmatrix}
        e^{-i\lambda x}\\0 \end{pmatrix}; \hspace{1cm}\boldsymbol{\phi}^{(1)} \xrightarrow[x \to \infty]{}\begin{pmatrix}
        a(\lambda)\,e^{-i\lambda x}\\   b(\lambda)\,e^{i\lambda x}\end{pmatrix} ,
\end{equation*}
The transmission $t(\lambda)$ and reflection $r(\lambda)$ coefficients are therefore:
\begin{equation*}
    t(\lambda)=\frac{1}{a(\lambda)}, \hspace{1cm} r(\lambda)=\frac{b(\lambda)}{a(\lambda)}.
\end{equation*}
The functions $a(\lambda) $ and $b(\lambda)$ determine the spectrum of the Zakharov-Shabat theory: the continuous spectrum is given by the value of $r(\lambda)$ computed for $\lambda\in\mathbb{R}$, while the discrete spectrum consists of the zeros of the coefficient $a(\lambda)$ in the upper complex half-plane $\mathbb{C}^+=\{\lambda\in\mathbb{C},Im(\lambda)>0\}$:
\begin{equation*}
a(\lambda_j)=0,\hspace{.5cm} \lambda_j=\zeta_j+i\eta_j, \hspace{.5cm}(\zeta_j,\eta_j)\in(\mathbb{R},\mathbb{R}^+).
\end{equation*}
 The spectrum $\cal S$ is therefore defined as the union of the continuous and discrete spectra:\\
\begin{equation*}
    \mathcal{S}=\{r(\lambda),\lambda\in\mathbb{R}\}\cup\bigg\{\lambda_m\in\mathbb{C}^+;\frac{b(\lambda)}{a'(\lambda)}\bigg|_{\lambda=\lambda_m}\bigg\}_{m=1}^N\;.
\end{equation*}
\subsubsection*{Tridiagonal form of the scattering functions}
Let us denote the elements of the transfer matrix defined in \eqref{transfer_M} as:
\begin{equation*}\label{T_j}
   T_j = \begin{pmatrix} a_j & \tilde b_j \\[1mm]  b_j & \tilde a_j \end{pmatrix},
\end{equation*}

where we use the shorthand notation for:
\begin{equation}
\begin{split}
&a_j = a(\lambda,|E^{(j)}|)=\cos(\Delta_j\mu_j)-i\lambda\frac{\sin(\Delta_j\mu_j)}{\mu_j},\\
& \tilde b_j = b(\lambda,E^{(j)})=i\frac{E^{(j)}}{\mu_j}\sin(\Delta_j\mu_j),\\
&  \tilde a_j = a(-\lambda,|E^{(j)}|) ,\hspace{.5cm}    b_j =  b(\lambda,\overline{E}^{(j)}).
    \end{split}
\end{equation}
with $\mu_j=\sqrt{\lambda^2+|E^{(j)}|^2}$.
It is worth noting that transfer matrices are in general non commutative, since:
\begin{equation}\label{commutator}
\begin{split}
[T_i&,T_j]=
\frac{\sin(\Delta_i\mu_i)\sin(\Delta_j\mu_j)}{\mu_i\mu_j}\times\\
&\times
\begin{pmatrix}
\overline{E}^{(i)}E^{(j)}-\overline{E}^{(j)}E^{(i)}
&
-2\lambda(E^{(i)}-E^{(j)})\\
2\lambda(\overline{E}^{(i)}-\overline{E}^{(j)})
&
-(\overline{E}^{(i)}E^{(j)}-\overline{E}^{(j)}E^{(i)})
\end{pmatrix}
\end{split}
\end{equation}
and consequences of that are discussed in the main section of this work.
It is convenient to introduce the modified Jost function $(p,q)^T$
\begin{equation}\label{Modified_Jost}
   \begin{pmatrix}
        p\\
        q
    \end{pmatrix}= e^{i\lambda x}\boldsymbol{\phi}^{(1)}  
\end{equation}
We then introduce the notation $(p_n,q_n)^T$ to denote the modified Jost function \eqref{Modified_Jost}, propagated beyond the n-th token:
\begin{equation*}
  \begin{pmatrix}
        p_n\\
        q_n
    \end{pmatrix}  =T_n \cdot T_{n-1}  \dots\cdot  T_1 \begin{pmatrix}
        p_0\\
        q_0
    \end{pmatrix}
\end{equation*}
Where the starting point\\
\begin{equation*}
\begin{pmatrix}
    p_0\\q_0\end{pmatrix}=\begin{pmatrix}
    1\\0\end{pmatrix},
\end{equation*}
represents the asymptotic state $x\to-\infty$. The effect of the single transfer matrix element $T_j$ determines the recurrence relation:
\begin{equation*}
    \begin{pmatrix}
        p_{n}\\
        q_{n}
    \end{pmatrix}=\begin{pmatrix} a_n & \tilde b_n \\[1mm] b_n & \tilde a_n \end{pmatrix}\cdot\begin{pmatrix}
        p_{n-1}\\
        q_{n-1}
    \end{pmatrix}, \hspace{1cm} n\ge 1
\end{equation*}
or explicitly:
    \begin{equation} \label{eq:coupled}
    \begin{split}
        &p_n = a_n\,p_{n-1} + \tilde b_n\,q_{n-1},\\
       & q_n =  b_n\,p_{n-1} + \tilde a_n\,q_{n-1}.
        \end{split}
    \end{equation}
   If we eliminate $q_{n-1}$ from the first equation of \eqref{eq:coupled} we get:
    \begin{equation*}
        q_{n-1} = \frac{p_n - a_n\,p_{n-1}}{\tilde b_n},
    \end{equation*}
 which, substituted into the second equation, gives
    the three-term recurrence of $p_n$ with variable coefficients:
    \begin{equation*}
        p_{n+1}= p_n\alpha_n- p_{n-1}\gamma_n
    \end{equation*}
    where:
{ \begin{equation*}
\begin{split}
\alpha_n = a_{n+1} + \tilde a_{n}\frac{\tilde b_{n+1}}{\tilde b_{n}}; \hspace{1cm}\gamma_n =\frac{\tilde b_{n+1}}{\tilde b_{n}},
\end{split}
    \end{equation*}}
where we have used that $\det(T_n)=a_n\tilde a_{n}-b_n\tilde b_{n}=1$. Given the initial conditions: 
\begin{equation*}
    p_{0}=1,\hspace{1cm} p_{1}=a_1,
\end{equation*}\\
the $p_{N}$ element of the ordered product can be written as the determinant 
of an $N\times N$ tridiagonal matrix\cite{Kittappa}:
\begin{equation*}
p_N(\lambda)=\det
    \begin{pmatrix}
       a_1 & 1        &          &          &          \\
        \gamma_1 & \alpha_1 & 1        &          &          \\
                 & \gamma_2 & \alpha_2 & 1        &          \\
                 &          & \ddots   & \ddots   & \ddots   \\
                 &          &          & \gamma_{N-1} & \alpha_{N-1}
    \end{pmatrix},
\end{equation*}
with $\Delta=\sum^N_j\Delta_j$. The scattering function is finally:
\begin{equation*}
      a(\lambda)=e^{i\lambda\Delta}p_N(\lambda).
\end{equation*}
 The simplest examples ($ N\le 4$) are listed below.
 \begin{itemize}
     \item N=1
 \begin{equation*}
 a(\lambda)=e^{i\Delta_1\lambda}a_1
 \end{equation*}
  \item N=2
 \begin{equation}\label{N_2}
 a(\lambda)=e^{i(\Delta_1+\Delta_2)\lambda}(a_1a_2+ b_1 \tilde{b}_2)
 \end{equation}
  \item N=3
 \begin{equation}\label{N_3}
 \begin{split}
 a(\lambda)=&e^{i(\Delta_1+\Delta_2+\Delta_3)\lambda}\times\\
 &\times(a_1a_2a_3+{b}_1\tilde b_2a_3+ {b}_1\tilde b_3\tilde{a}_2+{b}_2\tilde b_3a_1)
 \end{split}
 \end{equation}
  \item N=4
 \begin{equation}\label{N_quattro}
 \begin{split}
 a(\lambda)&=e^{i(\Delta_1+\Delta_2+\Delta_3+\Delta_4)\lambda}\big(a_1a_2a_3a_4+{b}_1\tilde b_2a_3a_4+\\
 &+{b}_1\tilde b_3\tilde{a}_2a_4+{b}_1\tilde b_4\tilde{a}_2\tilde{a}_3+{b}_2 \tilde b_3a_1a_4+\\
&+{b}_2\tilde b_4a_1\tilde{a}_3+{b}_3\tilde b_4a_1a_2+{b}_1\tilde b_2{b}_3\tilde b_4\big)
 \end{split}
 \end{equation}
 \end{itemize}
Note that the coefficients $b_j$ and $\tilde b_j$ are the only ones carrying the information of the spin $\sigma_j$, since the $a_j$ depend on the fields $E^{(j)}$ through its absolute value; therefore only the products between the $b_j$ or the $\tilde{b}_j$ represent the spin interaction elements. The formulas \eqref{N_2} and \eqref{N_3} show that  cases N=2,3 present only pairwise interactions, hence the first genuinely higher-order term is found at N=4, due to the presence of ${b}_1\tilde b_2{b}_3\tilde b_4$ in \eqref{N_quattro}, which represents the 4-spin interaction. This is the case used for computational purposes in the \textit{Discrete Spectrum Computing} section.
\end{document}